\documentclass{aa}

\usepackage{graphicx}
\usepackage{txfonts}
\usepackage{slashbox}
\usepackage{mhchem}
\usepackage{xcolor}
\usepackage[normalem]{ulem}

\definecolor{deepmagenta}{rgb}{0.8, 0.0, 0.8}

\usepackage{afterpage}

\usepackage[colorlinks=true,linkcolor=blue,citecolor=blue,urlcolor=blue]{hyperref}

\newcommand{\cair} {
  \ion{Ca}{ii}~8542~\AA\:}
  
\newcommand{\halpha} {
  \ce{H\alpha}}

\newcommand{\caK} {
  \ion{Ca}{ii}~K\:}
\begin{document}

   \title{Inference of the chromospheric magnetic field configuration of solar plage using the \cair line}

   \author{A.G.M. Pietrow\inst{1}
   \and D. Kiselman \inst{1}
   \and J. de la Cruz Rodr\'iguez\inst{1}
   \and C. J. D\'{i}az Baso\inst{1}
   \and A. Pastor Yabar\inst{1}
   \and R. Yadav\inst{1}
          }

   \institute{Institute for Solar Physics, Dept. of Astronomy, Stockholm University, Albanova University Centre, SE-106 91 Stockholm, Sweden\\
              \email{alex.pietrow@astro.su.se}
             }

   \date{Received June 26, 2020; accepted September 22, 2020}

\abstract{It has so far proven impossible to reproduce all aspects of the solar plage chromosphere in quasi-realistic numerical models. The magnetic field configuration in the lower atmosphere is one of the few free parameters in such simulations. The literature only offers proxy-based estimates of the field strength, as it is difficult to obtain observational constraints in this region. Sufficiently sensitive spectro-polarimetric measurements require a high signal-to-noise ratio, spectral resolution, and cadence, which are at the limit of current capabilities.}
{We use critically sampled spectro-polarimetric observations of the \cair line obtained with the CRISP instrument of the Swedish 1-m Solar Telescope to study the strength and inclination of the chromospheric magnetic field of a plage region. This will provide direct physics-based estimates of these values, which could aid modelers to put constraints on plage models.}
{We increased the signal-to-noise ratio of the data by applying several methods including deep learning and PCA. We estimated the noise level to be $1\cdot10^{-3} I_c$. We then used STiC, a non-local thermodynamic equilibrium (NLTE) inversion code to infer the atmospheric structure and magnetic field pixel by pixel.}
{We are able to infer the magnetic field strength and inclination for a plage region and for fibrils in the surrounding canopy. In the plage we report an absolute field strength of $|B| =440 \pm 90$~G, with an inclination of $10^\circ \pm 16^\circ$ with respect to the local vertical. This value for $|B|$ is roughly double of what was reported previously, while the inclination matches previous studies done in the photosphere. In the fibrillar region we found $|B| = 300 \pm 50$~G, with an inclination of $50^\circ \pm 13^\circ$.}{}


   \maketitle
%

\section{Introduction}\label{sect:1}

Observationally derived chromospheric magnetic field vectors are a vital part of our attempt to understand solar magneto-hydrodynamical processes, as is shown by the continuous effort that has been made in this direction over the last three decades \citep[e.g.,][]{bernasconi94,sami96,2000ApJ...544.1141S,2005A&A...436..325L, 2015SoPh..290.1607S,2016ApJ...825..119M,sara19,rahul19}.
Understanding these processes is of the utmost importance in also understanding the energy balance, atmospheric stratification, and dynamics of the solar chromosphere.

Plage regions, first observed in the chromosphere by \citet{Lockyer1869} and named by \citet{deslandres1893}, are classically defined as bright regions observed in \halpha\ and other chromospheric lines. Nowadays, many authors identify plage regions by only looking at the photospheric magnetic field concentrations, regardless of the \halpha\ intensity that is associated with those regions. In this paper we restrict the plage designation to regions where the magnetic field in the photosphere is confined in the intergranular lanes and forms a magnetic canopy in the chromosphere that is hot and bright in most chromospheric diagnostics 
such as the \ion{Ca}{ii} H\&K lines, the \ion{Ca}{ii}~infrared triplet lines, \ion{Mg}{ii}~h\&k, and \halpha. This leaves out superpenumbra, pores, or elongated fibrillar structures. Our usage is in line with that of \citet{Chintzoglou2020}.

Plage regions are important structures that act as the footpoints of coronal loops and the origins of fibrils, making them an important interface for coronal heating \citep{Reardon2009,Carlsson2019,2018A&A...615L...9C,Yadav20}. They have been extensively studied in the photosphere, leading to the view that the magnetic field in the lower layers is concentrated in the intergranular spaces \citep[e.g.,][]{buente93}. The inclination of the field in the photosphere is found to be mostly vertical, with average values slightly above $10^\circ$ from the local vertical \citep[e.g.,][]{bernasconi94, Sanchez94, MartinezPillet1997}. The flux tubes spread rapidly with height due to the exponentially decreasing gas pressure, eventually merging and filling almost the entire atmosphere. This leads to a decrease of the magnetic field strength with respect to height \citep[][and references therein]{sami92}. Perhaps the most clear detection of the magnetic field canopy effect from photospheric observations are the studies by \citet{Sanchez94} and \citet{2015A&A...576A..27B}, who detected the expansion of magnetic elements as a function of height using inversions.

However, due to the relatively low Land\'{e} factors and broad profiles of lines that are useful as chromospheric diagnostics, it is much more challenging to recover magnetic fields at this height when compared to the photosphere. Therefore, the field strength and the magnetic topology in plage are the subjects of an ongoing discussion. \citet{Carlsson2019} cite a canonical value of $|B|$ = 200~G. \citet{asensio2017} report transverse magnetic fields\footnote{We define the transverse magnetic field in our line of sight as $B'_\perp$ and the vertical magnetic field in the same reference frame as $B'_{||}$. Their unprimed counterparts represent the same quantities with respect to the solar surface.} with a median value of $B'_\perp \approx \:$ 60~G. This result is based on a Bayesian hierarchical model of the linear polarization of the data whose transverse magnetic field distribution has a tail going past 200~G. However, it is important to remark that the plage used in this paper is a young, flux-emerging region where the measurements were performed over elongated fibrillar structures, and therefore they do not fit our definition of plage.
However, despite this seeming agreement for the magnetic field strength, it has so far proven impossible to reproduce a plage chromosphere in quasi-realistic numerical models like Bifrost \citep{Gudiksen_2011}, which instead create an atmosphere with a calm chromosphere and cold corona, more resembling a coronal hole \citep{Carlsson2019}.

In recent years, significant progress has been made in the development of codes capable of performing non-local thermodynamic equilibrium (non-LTE) inversions of spectral lines in the chromosphere \citep{2008ApJ...683..542A,socas2015,Jaime16,millic18}. These inversions aim to provide a direct physics-based estimate of key parameters by producing a spatially resolved model atmosphere consistent with the observations, rather than proxy-based estimates that were the norm before. In this paper we perform non-LTE inversions to get a direct physics-based estimate of the magnetic field in plage, as well as other key parameters like temperature, line-of-sight velocity, and microturbulence. This is achieved by using polarimetric data with a high spatial, spectral, and temporal resolution in \cair as our input. In order to fully reconstruct the magnetic field vector, we use a combination of several novel and standard post-reduction techniques to improve the otherwise low signal-to-noise ratio of our Stokes $Q$ and $U$ measurements. 

In Sect.~\ref{about8542} we discuss the diagnostic potential of the \cair line and the limitations that come with it. In Sect.~\ref{observations} we focus on our observations, the reduction, and post-reduction steps. In Sect.~\ref{aboutstic} we discuss the STockholm Inversion Code (STiC) and the additional steps that we applied to aid the inversions. Response functions and uncertainties are discussed in Sect.~\ref{abouterror}. The results from our inversions are shown and discussed in Sect.~\ref{sec:res} and our final conclusions can be found in Sect.~\ref{conclusions}.

\section{The \ion{Ca}{ii}~8542~\AA\ line in plage regions}\label{about8542}
\begin{figure}[!ht]
   \centering
   \includegraphics[width=\columnwidth]{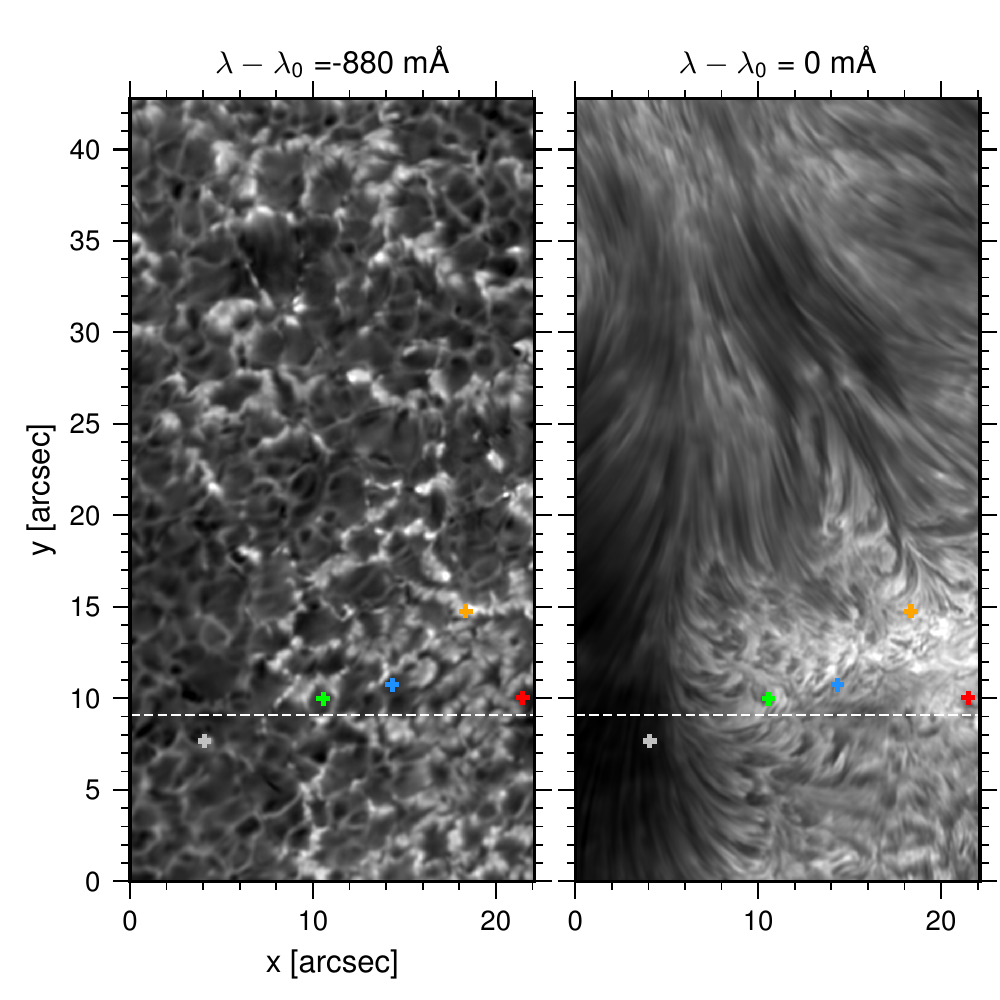}
   \includegraphics[width=\columnwidth]{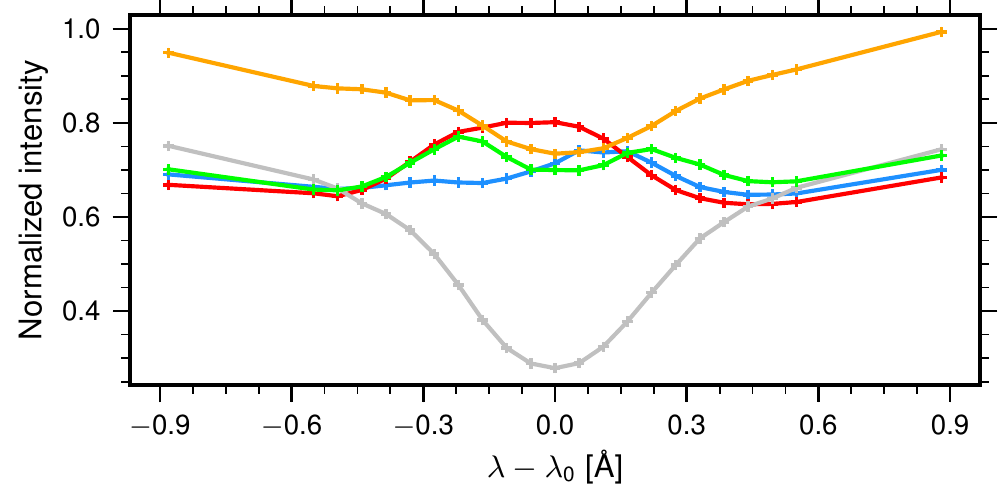}
   \includegraphics[width=\columnwidth]{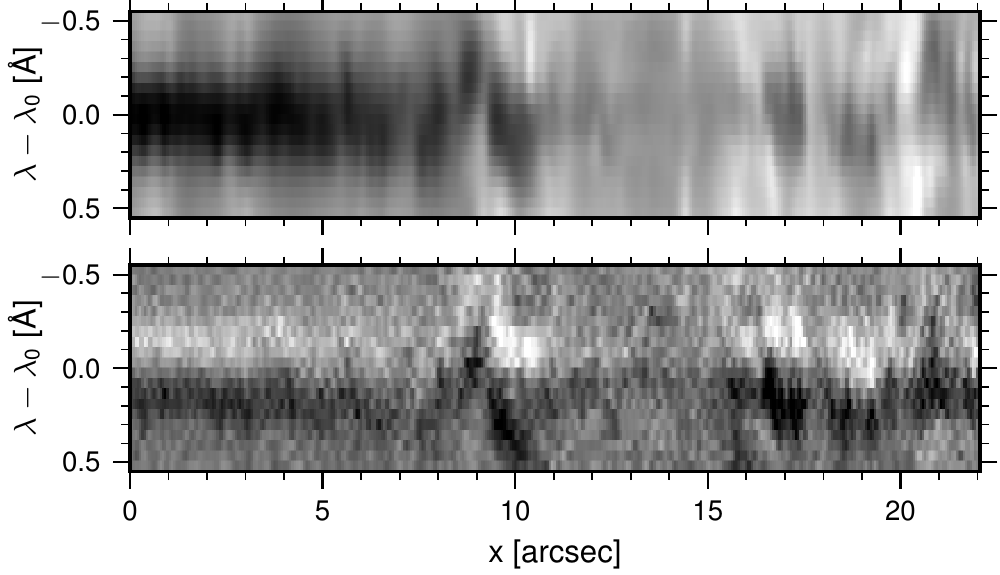}
   \caption{Plage spectral signatures in the \ion{Ca}{ii}~8542~\AA\ line. Top: Monochromatic images acquired in the wing ($\Delta\lambda=-880$)~m\AA\ and in the core of the line. Middle: Example plage spectra in Stokes~$I$ extracted from the locations indicated in the upper panels with the same color coding. For comparison, we have also included a quiet-Sun profile colored in gray. Bottom: Stokes~$I_\lambda$ (upper) and $V_\lambda/I_\lambda$ spectra (lower) along the slit indicated in the top panels of the figure.}
   \label{fig:line}
\end{figure}

\begin{figure*}[!htb]
 \includegraphics[width=\linewidth]{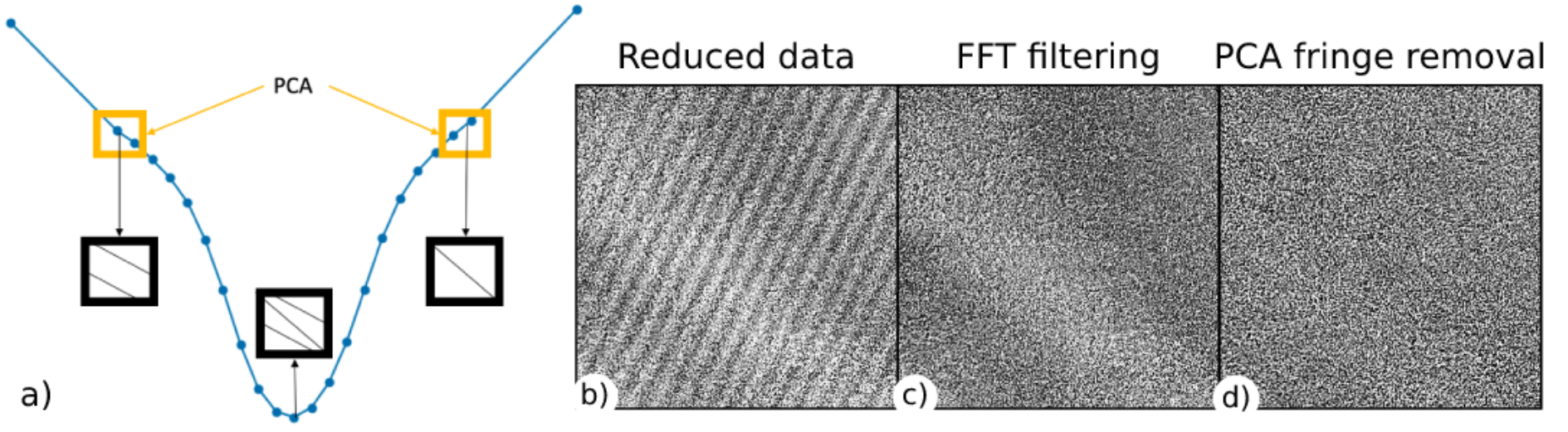}

\caption{Fringe removal from reduced data. a) A cartoon of the low-frequency fringe extraction process. Each wavelength point contains a linear combination of the fringes found in the left and right wing points, which are denoted by several lines. These fringes on either side could be extracted from the wings applying principle component analysis (PCA) to the outer most non-continuum wing points. A linear combination of these two fringes are then subtracted from the full spectrum. b) A Stokes $U$ line wing image showing both the high- and low-frequency fringes. c) The same image after the high-frequency fringes have been removed with FFT (
Fast Fourier transform) filtering. d) The same image after the low-frequency fringe was removed using our PCA fringe removal technique.}\label{fig:PCA}
\end{figure*}

The \cair line has become one of the foremost chromospheric diagnostics for temperatures, line-of-sight velocities, and the magnetic field vector \citep[see, e.g.,][]{2000Sci...288.1396S,2001ApJ...552..871L,2007ApJ...670..885P,2012A&A...543A..34D,2015ApJ...810..145D,2017MNRAS.472..727Q,2017ApJ...845..102H,2018ApJ...866...89C,2019A&A...623A.178D}. Although the atomic level populations must be modeled under the assumption of statistical equilibrium \citep{2011A&A...528A...1W}, a relatively simple six-level atom can be used to model the intensities of the \ion{Ca}{ii}~H\&K lines as well as the infrared triplet lines. The profile of this line can be calculated by assuming a complete redistribution of scattered photons \citep{1989A&A...213..360U,2017A&A...597A..46S}. In active regions far from the limb, the imprint of atomic polarization can be assumed to be much lower than Zeeman-induced polarization \citep{2010ApJ...722.1416M,2016ApJ...826L..10S}.

In plage targets, the \cair line shows peculiar line profiles that originate in a hot magnetic canopy that extends over a photosphere where the magnetic fields are tightly concentrated in the intergranular lanes \citep{jaime13}. The main feature of those profiles is the apparent absence of an absorption Gaussian core. Instead, the core of these profiles can appear completely flat, or it can show weak absorption and emission features that are modulated by the velocity field and a very shallow stratification of the source function \citep[][]{jaime13,Carlsson2015}. In the top panels of Fig.~\ref{fig:line} we show a subfield from our observations that includes plage and some quiet-Sun next to it. We have hand-picked some locations in the plage region and plotted the corresponding profiles in the middle panel. Compared to the quiet-Sun profile (gray), the plage profiles show a relatively raised core with a diversity of shapes and asymmetries. While the green, blue, and red profiles originate from a non-magnetic photosphere and a magnetic chromospheric canopy, the orange profile is located over a magnetic element that is also brighter in the photospheric wings. This line-core effect is therefore not present, but instead the whole profile is brighter than that of the quiet-Sun.

The peculiar shape of the line sets certain limitations on the diagnostic potential of the \cair line in plage targets:
\begin{enumerate}
    \item The relatively flat core of the line makes it very hard for an inversion code to constrain the Doppler width of the line and, therefore, to derive accurate values of microturbulent motions. The explanation for the relatively large values of microturbulent velocity that are required to explain some chromospheric lines \citep{1974SoPh...39...49S, Carlsson2015,2015ApJ...799L..12D} remains an open question that can hardly be tackled with observations of this line in plage. 
    \item The amplitude of Stokes~$Q$, $U,$ and $V$ is modulated by the magnetic field strength (in the weak-field regime) and by the gradient of the source function. When the latter is large, the amplitude of the Stokes parameters becomes large. This dependence is explicit even in simple analytical solutions of the radiative transfer equation such as the Milne-Eddington one \citep[see, e.g., ][]{2007A&A...462.1137O}. In those locations where the source-function gradient becomes very shallow, the line core becomes particularly flat and therefore the amplitudes of the Stokes parameters are small. An example of this effect is visible in the lower panels of Fig.~\ref{fig:line} around $x\approx14\arcsec$. However, whenever the source function is not strictly zero, the amplitudes of the Stokes parameters seem to be sufficiently high even within plage patches ($x\ge 16\arcsec$ along the slit in Fig.~\ref{fig:line}).
    \item Due to the relatively flat line-core shape, these profiles do not always provide complete information about line-of-sight velocities or about line-of-sight velocity gradients in the chromospheric part of the line.
\end{enumerate}

These limitations can affect the sensitivity of the line to a given physical parameter across the field-of-view (FOV) and they should be kept in mind in order to properly understand the inversion results that we present in Sect. \ref{sec:res}.
\afterpage{
\begin{figure*}[!htp]
   \centering
   \includegraphics[width=0.79\textwidth, trim=1.3cm 1.8cm 1.5cm 3.6cm,clip]{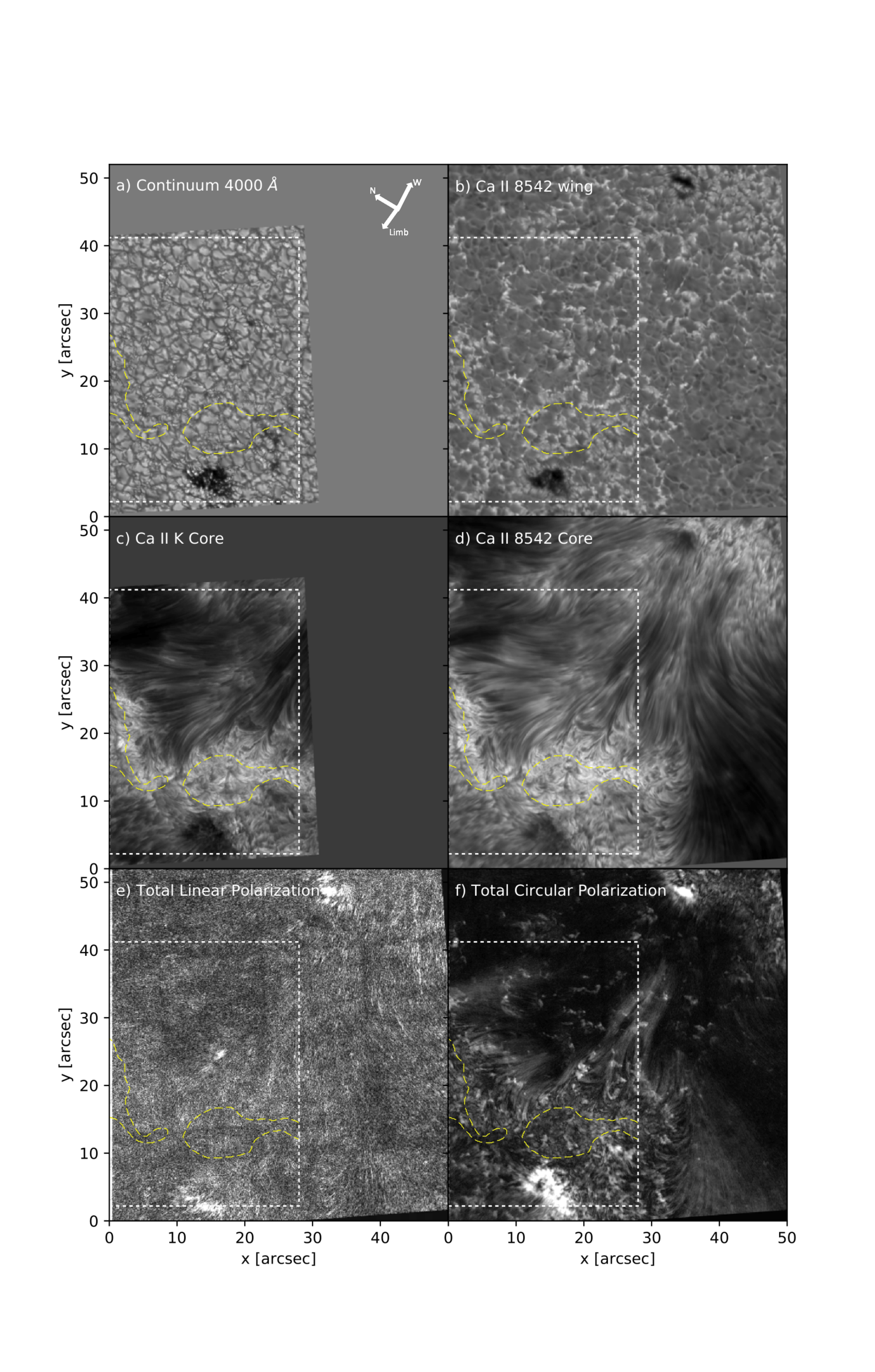}
   \caption{Overview of AR12713 taken on June 15th 2018 at 14:23 UT. Overlap between CHROMIS and CRISP is marked with a white rectangle. The plage areas P1 and P2 are marked with yellow contours. a) Continuum intensity 4000\,\AA. b) Ca II K core intensity. c) Ca II K at $\Delta\lambda$ = 0.26\,\AA. d) \cair line core intensity. e) Total linear polarization based on the wavelength average of Stokes $Q$ and $U$. f) Total circular polarization based on the wavelength average of Stokes $V$.}
   \label{fig:overview}
\end{figure*}

}
\section{Observations and data processing}\label{observations}
Region AR12713 was observed on June 15 2018, between 14:23 and 14:48 UT with the Swedish Solar 1-m Telescope \citep[SST,][]{Scharmer03}, using both the CRisp Imaging SpectroPolarimeter \citep[CRISP,][]{Scharmer08} and the CHROMospheric Imaging Spectrometer \citep[CHROMIS;][]{2017psio.confE..85S} instruments simultaneously. The region was centred around the heliocentric coordinates  (x,y) = ($-557\arcsec$, $80\arcsec$), which translates to an observing angle of 37$^\circ$ ($\mu$ = 0.80). This region was selected for this reason, as an observing angle close to 45$^\circ$ of a mostly vertical field ensured a signal in both the horizontal and vertical line-of-sight components of the magnetic field. The measurements were acquired using a high-cadence program for both instruments. We optimized the cadence, given the limitations of a scanning instrument, to reach a compromise between signal-to-noise ratio (S/N) and temporal resolution. This has allowed us to apply several of the methods below, as well as to bin consecutive scans for a higher S/N. This high-cadence program only allowed for one line to be observed. In our case this was \cair for CRISP and \caK for CHROMIS.

For CRISP the cadence was 9.6~s and the observing sequence consisted of 23 wavelength positions in the \cair line, ranging from -0.88 to +0.88~$\AA$ relative to line center, with an equidistant spacing of 0.055~$\AA$ for the inner 21 points and one doubly spaced point on each end. The CRISP pixel scale is $0.058\arcsec$.
For CHROMIS the cadence of the observing sequence was 9.0~s with 28 wavelength positions in the \caK line, ranging from -1.51 to +1.51~$\AA$ relative to line center. These observations have an equidistant spacing of 0.065~$\AA$ for the inner 21 points, as well as two 0.262~$\AA$ steps on each side, a final wing point 0.328~$\AA$ after that, and one single continuum point at 4000.08~$\AA$.  The CHROMIS pixel scale is $0.0375\arcsec$. Both data sets were reduced using the CRISPRED and CHROMISRED pipelines (Now combined into SSTRED) as described by \citet{2015A&A...573A..40D} and \citet{2018arXiv180403030L}, which make use of Multi-Object Multi-Frame Blind Deconvolution \citep[MOMFBD;][]{vanNoort05,2002SPIE.4792..146L}.

In this paper we mainly focus on the \cair data, only using the \caK profile for an initial line-of-sight velocity estimate. Because of this, the CHROMIS data were binned down to the CRISP pixel scale and aligned to these observations, as opposed to up-sampling CRISP as is usually the case. In order to improve the S/N of the resulting hypercube (t, x, y, $\lambda$, s), we applied five separate post-processing techniques to the data, which we will cover in order of application. 

\subsection{Fourier filtering}
In our observations we could distinguish a high-frequency fringe diagonally across the image, and a low-frequency fringe perpendicular to it. The high-frequency fringes had the strongest amplitude in Stokes $U$, at $1.5\cdot 10^{-3}\,I_c$. Stokes $V$ had half that, Stokes $Q$ had 13\%, and Stokes $I$ only 5\%. While negligible in Stokes $I$ and $Q$, these fringes  needed to be removed from the Stokes $U$ and $V$ maps before any inferences could be made from them. The fringes are stationary on the detector but seem to move in the final data cube due to image derotation being applied to the time series. Such fringes can be removed in the frequency domain by using a two-dimensional Fourier transform of the image \citep{lim1990two}. In this work the removal was done by applying a simple mask to the frequencies that corresponded to the fringe pattern, after which the image was transformed back into the spatial domain. We then constructed the wide- and narrow-band cubes out of these filtered images with standard routines. This method is similar to the procedure used for removing fringes in observations made by the Sunrise telescope \citep{MartnezPillet2010}; a similar approach is now part of the CRISPEX pipeline. In Fig.~\ref{fig:PCA} b and c we can see the effects of this method.

The low-frequency fringes could not be removed in the same way, since masking at lower frequency would lead to significant loss of data. Instead we removed these fringes by applying PCA at a later step.

\subsection{Improving S/N with a convolutional neural network}
Recently, several studies have shown how convolutional neural networks \cite[][]{lecun1995} could help to reduce the noise in observations, showing a better performance than classical methods. These neural networks exploit the idea of using the presence of spatial correlation to predict the value of a pixel from the value of other pixels. In ideal examples, artificial noise can be used to create a sample where the neural network can learn from it. However, the characterization of noise in solar applications is not always trivial and the noise is spatially correlated due to the application of MOMFBD. Using high-cadence observations, a neural network can be trained to infer the contribution of the noise from the temporal redundancy of the signals themselves. This technique  \citep[also known as Noise2Noise;][]{lehtinen2018} has specifically been explored on CRISP spectro-polarimetric data by \citet{Carlos19} and therefore could be directly applied to our observations in Stokes $Q$, $U,$ and $V$.
For details on the implementation of the network, we direct the reader to the original paper and the project GitHub\footnote{\url{https://github.com/cdiazbas/denoiser}.}. This step significantly lowered the noise floor of our observations by a factor of four.

\subsection{Binning and stacking}
After removing the high-amplitude, high-frequency fringes from the data and denoising it with the neural network, we binned and stacked the resulting hypercube in order to improve the S/N. Despite the high cadence of the observations, we could not stack more than two scans in the temporal domain without ending up with non-physical mixed profiles. This maximum was established by comparing the Stokes $I$ profiles of the resulting average to those of the individual scans. As long as these were the same, we could stack images. The same could be done with spatial binning, where the maximum was found to be 2x2 binning. We stacked two frames and binned the image down two times. 

\subsection{Fringe removal and S/N improvement using principle component analysis}
Principal component analysis \citep[PCA,][]{Pearson1901} is a linear orthogonal transformation that converts a set of possibly correlated variables into a set of principal components (PC), which are linearly uncorrelated variables, sorted by means of variance. In other words, one could imagine the method as fitting a p-dimensional ellipsoid to the data, with each axis of the ellipsoid representing a principle component and the length of each radius being defined by its variance. A small radius corresponds to a small variance, and the general fit will hardly be affected if this axis is omitted, meaning that we only lose a small amount of information.
In this work we used PCA in two ways: to remove the low-frequency fringes from our data and to improve the S/N. In the former case we used the method together with the assumption that the fringe pattern at each wavelength point is a linear combination of the pattern found in the wings of the line. On top of that we assumed that there was no signal in the outer wing points of Stokes $Q$ and $U$, so that the intensities in these points consist of the fringe and background (see Fig. \ref{fig:PCA}a). In order to separate the fringes from the background, we applied PCA to the two outermost non-continuum wing points on both sides of the line. From each wavelength position we then subtracted an approximation of the fringe pattern  obtained by taking the linear combination of the two outermost fringe patterns with  

$$ f_{\lambda} = f_{\rm red}\cdot(1 - \lambda/n) + f_{\rm blue}\cdot(\lambda/n), $$ 
where $f_\lambda$ is the fringe intensity at wavelength position $\lambda$, $f_{\rm red}$ and $f_{\rm blue}$ are the extracted fringe intensities from the red and blue wing respectively, and $n$ is the number of data points in the observation (see Fig.~\ref{fig:PCA}d,e).

After applying the PCA fringe extraction, we applied another PCA routine to further improve the S/N of the data. Following the methods described, for example, in \cite{Casini2012}, \cite{Ng2017}, \cite{casinili18}, and \cite{MartnezGonzlez2008}, we applied PCA along the wavelength axis of the 3D cubes of the individual Stokes parameters. The resulting set of 23 principal components contained most of the information in the first 15 components, with the background noise composing the most significant part of the final eight. Therefore removing these principal components and reconstructing the data with the remaining set will improve our S/N ratio, while only losing a minimal amount of data. We applied this method to Stokes $Q$ and $U$. It was not necessary to apply it to Stokes $I$ and $V$ due to their relatively higher S/N. After applying the methods discussed above, we estimated that the background noise value is $1\cdot10^{-3} I_c$ for calculating weights in the inversion. 

\subsection{Wavelength and intensity calibration}
Finally we performed  an absolute wavelength and intensity calibration using the solar atlas by \cite{Neckel1984}. During these post reduction steps, we  extensively used the CRISPEX analysis tool \citep{Gregal12}, the CRISpy python package \citep{pietrow19}, and SOAImage DS9 \citep{2003DS9} for data visualization. 

\subsection{Resulting field of view}
In Fig.~\ref{fig:overview} we give an overview of the field of view (FoV) that was obtained after the reduction and post-reduction. We focus on a single temporally averaged scan taken at 14:23 UT, which was chosen for its good seeing. After binning in space and time the cadence is 19.2~s and the spatial scale is 0.116\arcsec/px. Solar north and the limb are marked in panel a.
The dashed rectangle indicates the FoV common between CRISP and CHROMIS and will be subject to data inversions.

In the full CRISP FoV we see two pores, one on the upper left side and one on the lower right side, best visible in panel b. In panel d we can see that both of these pores are surrounded by unipolar plage regions that are connected with a carpet of fibrils. These fibrils originate from the lower plage region and are moving up to, and around, the upper plage region. We have marked the plage within the white rectangle with a yellow contour by selecting the bright areas in panel c. We designate these two plage regions as P1 and P2. The regions were defined by selecting bright regions from the line-core image and then adjusting them by hand to exclude any fibrillar structure. The size of these regions is 5409 and 2561 pixels respectively.
The last two panels show the polarization maps for the total linear (e) and circular (f) polarization after post-reduction.

The magnetic field in the canopy is expected to be mostly horizontal with respect to the solar surface, while in the plage it is expected to be mostly vertical. However, due to the fact that we observe at a viewing angle of $37^\circ$ ($\mu=0.8$), we expect to see imprints of the plage and the canopy in both polarization maps. This largely seems to be the case.

\section{Data inversions}\label{aboutstic}
\subsection{Stockholm Inversion Code}
We performed an inversion  of the post-processed Stokes profiles for our observations with the STockholm inversion Code\footnote{\url{https://github.com/jaimedelacruz/stic}} \citep[STiC;][]{Jaime16,Jaime19}, a 
parallel non-LTE inversion code that utilises a modified version of the radiative-transfer code RH \citep{Uitenbroek01} to solve the atomic population densities by assuming statistical equilibrium and plane-parallel geometry. STiC uses a regularized version of the Levenberg-Marquardt algorithm \citep{levenberg44, marquardt63} to iteratively minimize the $\chi^2$ function between observed input data and synthetic spectra of one or more lines simultaneously. The code treats each pixel separately as a plane parallel atmosphere, fitting them independently of each other.

We inverted all four Stokes parameters of the \cair line simultaneously, also including a 4000~$\AA$ continuum point from the \caK observations. We could not include the \caK line data  due to the temporal averaging that has been used to increase the S/N of the linear polarization data. Therefore there is a relatively narrow range in optical depth where we have enough sensitivity to do inversions. As discussed in Sect.~\ref{about8542}, it is challenging to reconstruct a good velocity and microturbulence estimate based on \cair alone in plage. We attempted to create a better constraint on the velocity by introducing a line-of-sight velocity estimate from the available \caK data, which usually has a regular emission core in plage with a large amplitude.

\subsection{Initialization of the line-of-sight velocity}\label{initvelocity}
To get a better line-of-sight velocity estimate we used a method similar to that suggested by \cite{Skumanich02}. Here the authors show that a first order approximation of the velocity map can be made by reconstructing a red- and a blue-shifted profile with a basis composed of the first and second eigenprofiles of the intensity and fitting a linear relation to the resulting factors. This method works as long as the second eigenprofile is similar to the derivative of the first. Unfortunately, this was not the case for our data, due to the relatively small number of wavelength points. We instead chose to directly define a basis that is composed of the average intensity profile in the plage and its derivative. This basis was not fully orthonormal, but still gave a usable estimation of the velocity where the PCA basis breaks down completely. The chromospheric velocity estimate was obtained by using the line core and the photospheric estimate was made by using the outer wing points (see Fig.~\ref{fig:vlos} b and c).
These two maps were then combined into a velocity estimate for the atmosphere where the chromospheric velocities were used for $\log(\tau_{500}) < -3.5$ and the photospheric velocities for the lower atmosphere. (Hereafter we drop the 500 from $\log(\tau_{500})$
to simplify the notation.) An arctangent function was used to connect the two in a smooth way. The resulting input model can be seen in Fig.~\ref{fig:vlos}a.

\subsection{Inversion approach}
We initialized the first cycle of STiC by generating a FAL-C \citep{Avrett85,fontenla93} atmosphere interpolated over 54 depth points in the range of $\log(\tau) = [-7,1]$, which we combined with our initial velocity estimate from Sect.~\ref{initvelocity}. Moreover, we used the first three inversion cycles to obtain a model that approximates the atmosphere in Stokes $I$ and $V$. We then set an initial guess for the magnetic field in the form of the weak field approximation \citep{landi2004} for $B'_{||}$, 800~G for $B'_{\perp}$ , and 1.2 for $\chi$. The last three inversion cycles were then used to fit Stokes $Q$ and $U$. An overview of the cycles can be found in Table \ref{tab:invert}. We only inverted the part of the FoV that had overlap with the CHROMIS data, as this is the only part where we can get an estimate for the velocity and have information on the 4000~$\AA$ continuum. The latter helps to constrain the temperature reconstruction in the photosphere. As a final step we subtracted the CRISP cavity error map from the resulting velocity. We raised the weights of Stokes $Q$, $U,$ and $V$ to ensure that the code took these into account. The initial weight of $1\cdot10^{-3} I_c$ was divided by four for Stokes $V$ and ten for Stokes $Q$ an $U$.

\section{Response functions and uncertainties}\label{abouterror}
The \cair line is sensitive to a large range of $\log(\tau)$, with the line core being formed in the chromosphere and its wings reaching down to the photosphere. Additionally, several of the aforementioned effects will affect the quality of our inversions more at certain heights than others. For this reason it is beneficial to compute response functions (RF) of our parameters \citep{beckers75,innocenti77,1992SoPh..137....1S,1994A&A...283..129R,millic17}. A RF is the derivative of the emerging intensity as a function of a given physical parameter per $\delta\log(\tau)$. It also allows us to estimate how strongly the Stokes parameters react to a perturbation at a certain atmospheric height. The stronger the RF, the more sensitive the line is at a given $\log(\tau)$ for the perturbed quantity. The shape of the RF depends on the magnitude of the perturbation, which should be just above the numerical noise \citep{millic17}. We used the built-in RF mode in STiC to compute our response functions and used the same perturbation size as has been used for the inversions. After generating our RFs for our models, we found that the maximum response for both magnetic field parameters  lies between $\log(\tau) = [-5.5,-3.5]$. Outside this optical-depth range, the response is negligible.

The goodness of the fit can be quantified by computing the $\chi^2$ value between the observed and synthetic data. In our case we defined the function as
\begin{equation}
 \chi^2 = \frac{1}{4q} \sum\limits^3_{s=0}\sum\limits^q_{i=1} \left( \frac{I_s^{obs}(\lambda_i) - I^{syn}_s(\lambda_i;M) }{w_{s,i}} \right)^2,
 \label{eq:chi2}
\end{equation}
where $q$ is the number of wavelength points of the Stokes profiles, $I^{syn}$ are the Stokes profiles from a model $M$ while $I^{obs}$ are the observed Stokes profiles. The ratio of the weights factor and noise of the observed
Stokes profiles are denoted by $w_{s,i}$  .

Additionally we could use the calculated RFs to estimate the uncertainty in our resulting physical quantities. The latter was done following the method described in \citet{delToroIniesta2003}, with the exception that we only introduced perturbations at the locations of our inversion nodes directly. The uncertainty in a given physical quantity $p$ can be written as

\begin{equation}
\centering
    \sigma^2_p = \frac{2}{n} \frac{\sum\limits^3_{s=0}\sum\limits^q_{i=1} \left[ I_s^{obs}(\lambda_i) - I^{syn}_s(\lambda_i;M) \right]^2 w^2_{s,i} }{\sum\limits^3_{s=0}\sum\limits^q_{i=1} R^2_{p,s}(\lambda_i) w^2_{s,i}},
    \label{eqerr}
\end{equation}
where $n$ is the number of nodes used in the inversion in model $M$ and $q$ the amount of wavelength points in the observation and $R$  is the response function of a Stokes parameter  to  the  physical  quantity $p$. 

We picked one typical pixel in P2, F1, and F2 for which we calculated the uncertainties in all six fitted parameters using Eq. \ref{eqerr}. We displayed these profiles and their fit along with the retrieved atmospheric parameters and their uncertainties in Fig. \ref{fig:RF}. The three profiles have a respective $\chi^2$ value of 6, 9, and 5, which correspond to yellow and light green in the map. A good fit would have a $\chi^2$ of roughly 1 if the data is used without weights, but when we use weights to be more sensitive to weak amplitudes in the polarization, larger values will be obtained for a similar fit \citep{Milic20}.
The location of these representative pixels is shown in Fig. \ref{fig:bfield}. In Fig. \ref{fig:chi2}a we displayed the $\chi^2$ values of the entire FoV.

\begin{figure*}
   \centering
   \includegraphics[width=1\textwidth]{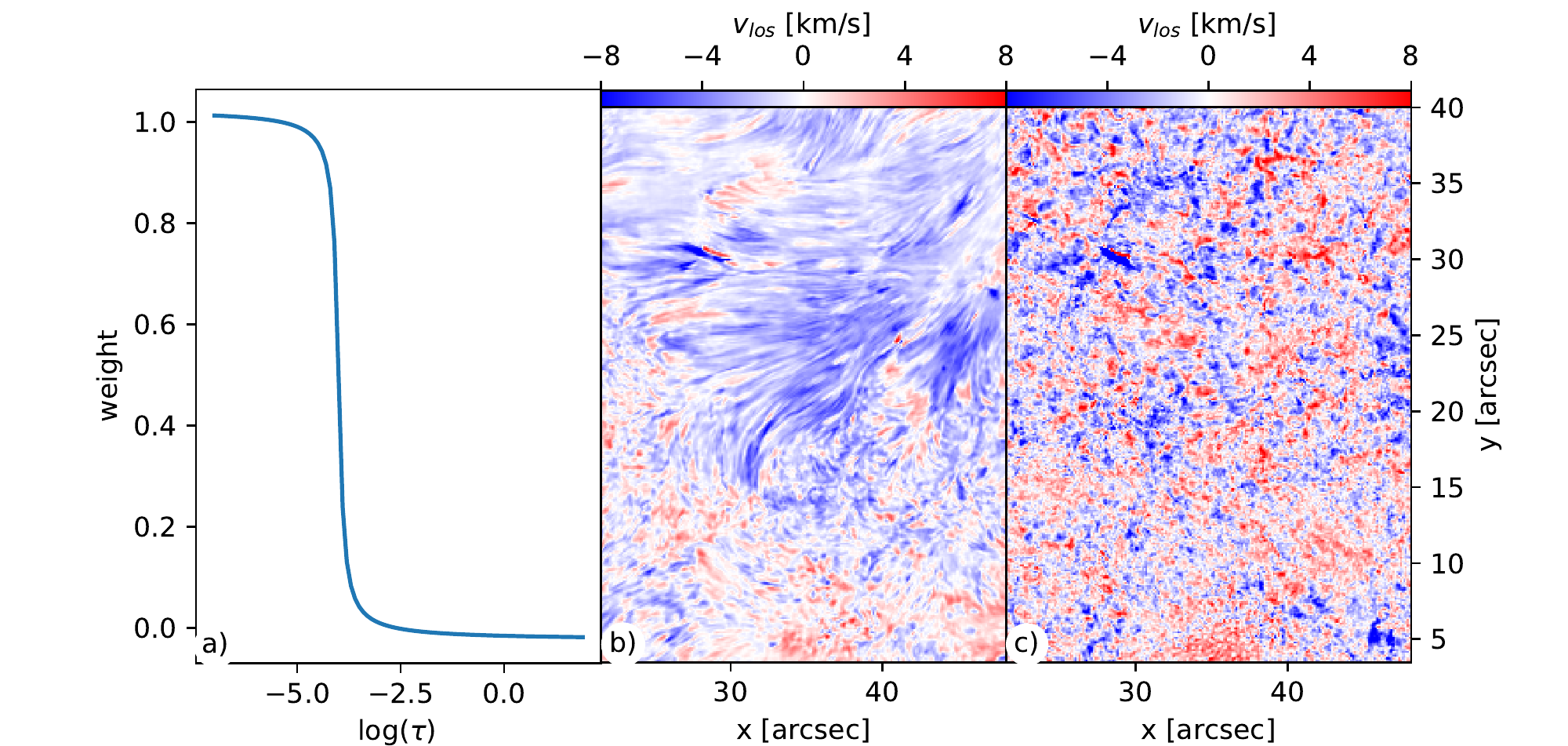}
   \caption{Simple two-component velocity model created by applying our linear velocity estimates based on the line core and line wings of \caK. We transition from the photospheric velocity estimate to the chromospheric velocity estimate at $\log(\tau) = -3.5$ by means of an arctangent function. This velocity model was used together with a FAL-C atmosphere as the initial guess for the inversion code. From left to right the scaling functions used to compose the velocity model (a), the chromospheric velocity estimate (b) and the photospheric velocity estimate (c).}
   \label{fig:vlos}
\end{figure*}

\begin{table}
\caption{\label{tab:invert} Overview of the number of nodes used per inversion cycle.}
\begin{center}
\begin{tabular}{|c||*{6}{c|}}\hline
\backslashbox{Parameter}{Run \#} &  1 &2 & 3& 4&5 &6 \\ \hline\hline
T &5&7&7&7&0&7\\\hline
$v_{los}$ &0&0&2&3&0&3\\\hline
$v_{turb}$ &0&0&1&1&0&1\\\hline
$B'_{||}$ &0&0&0&1&0&3\\\hline
$B'_{\perp}$ &0&0&0&1&1&1\\\hline
$\chi'$ &0&0&0&1&1&1\\\hline

\end{tabular}
\end{center}
      \small
       {\textbf{Note:} Columns specify the amount of nodes used per cycle per parameter. The resulting atmosphere was smoothed between each run, except for the fifth cycle where we only varied $B'_{\perp}$ in order to facilitate a better fit. The initial model was a standard FAL-C atmosphere with 54 points in the optical depth ranging from -7 to 1, together with our initial velocity estimate.}

\end{table}

\section{Results and discussion}\label{sec:res}
\begin{figure*}
   \centering
   \includegraphics[width=0.99\textwidth, trim=2.5cm 0cm 3cm 0cm,clip]{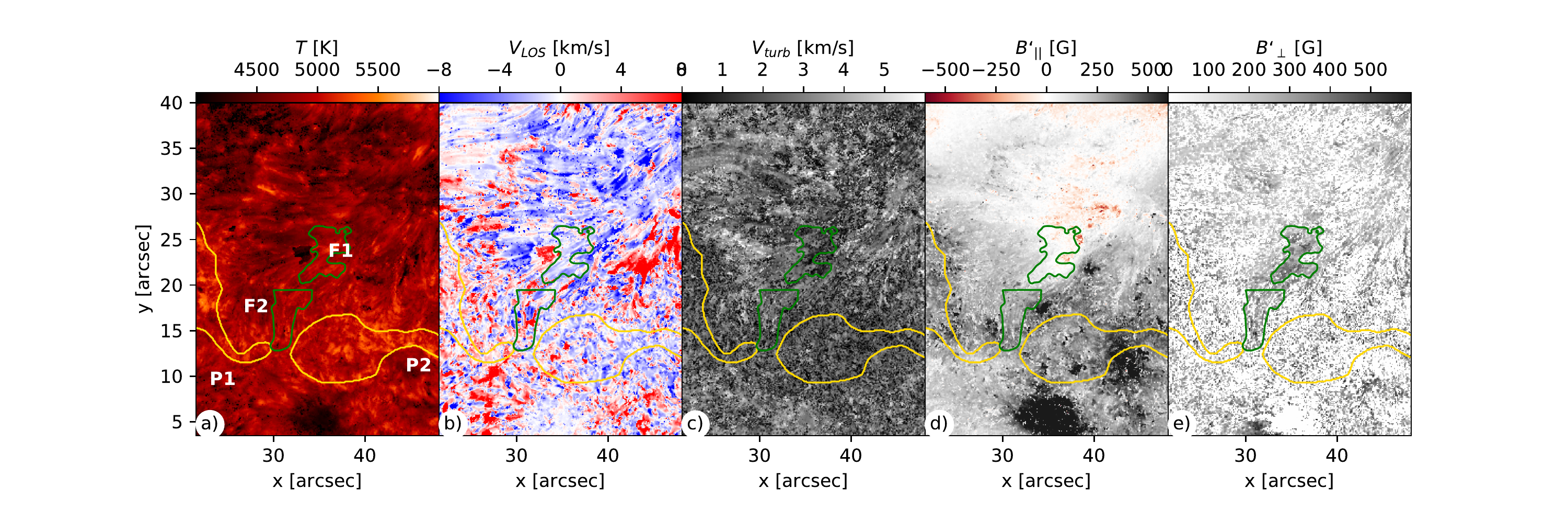}
   \caption{Inversion results after six reduction steps in STiC as shown in Table. \ref{tab:invert}. All panels are plotted at $\log(\tau) = -4.5$. Two plage regions P1 and P2 are marked in yellow. Inside the region covered with fibrils we mark two regions as F1 and F2. From left to right: Temperature (a),  line-of-sight velocity (b),  microturbulence (c), longitudinal magnetic field (d), and the transverse magnetic field (e).} 
   \label{fig:inversions}
\end{figure*}

In Fig.~\ref{fig:inversions}, we present the inversion results of the $37\arcsec \times 27\arcsec$ field of view that was marked with a white dotted line in Fig.~\ref{fig:overview}. Since the maximum sensitivity for $B'_{||}$ and $B'_{\perp}$ in our response functions lies between $\log(\tau) = [-5.5,-3.5]$, these results are displayed for a height of $\log{(\tau)} = -4.5$.  

Our chromospheric temperature map is largely consistent with the \cair line-core-intensity map in Fig.~\ref{fig:overview}d. The median temperature is around 4800~K in the fibrillar region, and around 5500~K, with peaks up to 6000~K, in plage. Our values are consistent with earlier reported plage temperatures at this height when taking into account the uncertainty of 430 K, obtained from Eq. \ref{eqerr} as shown in Fig. \ref{fig:RF}. \cite{2019A&A...623A.178D} report 5000 K and \cite{joao2020} report values between 6000 and 6500K. The temperature of the fibrillar region is also comparable to the values reported by \citet{sepideh20}.

The reconstruction of the line-of-sight velocity is not very accurate due to the raised core profiles of the \cair line in plage as discussed in Sect.~\ref{about8542}.  When we compare the estimate based on \caK (as seen in Fig.~\ref{fig:vlos}b) to the velocity from the \cair line, (see Fig. \ref{fig:inversions}b.) we can clearly see the limitations. The velocity in the canopy areas with fibrils is much smoother and better constrained than in the plage, where we retrieve a velocity field with a more patchy look.

Since the Doppler width constrains the turbulent velocity parameter in the inversions, there is low sensitivity to this parameter. Having at least two lines from species with a well-differentiated mass would certainly help to constrain the microturbulence, but was not the case in our observations. Therefore we have used a low-norm regularization being set to favor lower microturbulence values in the reconstruction. The latter resulted in microturbulence values systematically below 2~km~s$^{-1}$ in the plage region. As expected, these values were significantly lower than those reported by \citet{Jaime16}, \citet{ 2015ApJ...799L..12D}, \citet{Carlsson2015} and \citet{joao2020}, where high microturbulence values are required in order to fit \ion{Mg}{ii} h\&k profiles in plage. 
Outside the plage we regained some sensitivity to microturbulence, and reach values up to 6~km~s$^{-1}$, which is comparable to values derived from similar data inversions to those of \citet{sepideh20}.

\begin{figure*}
   \centering
   \includegraphics[width=\textwidth, trim=1cm 1.2cm 1cm 1cm,clip]{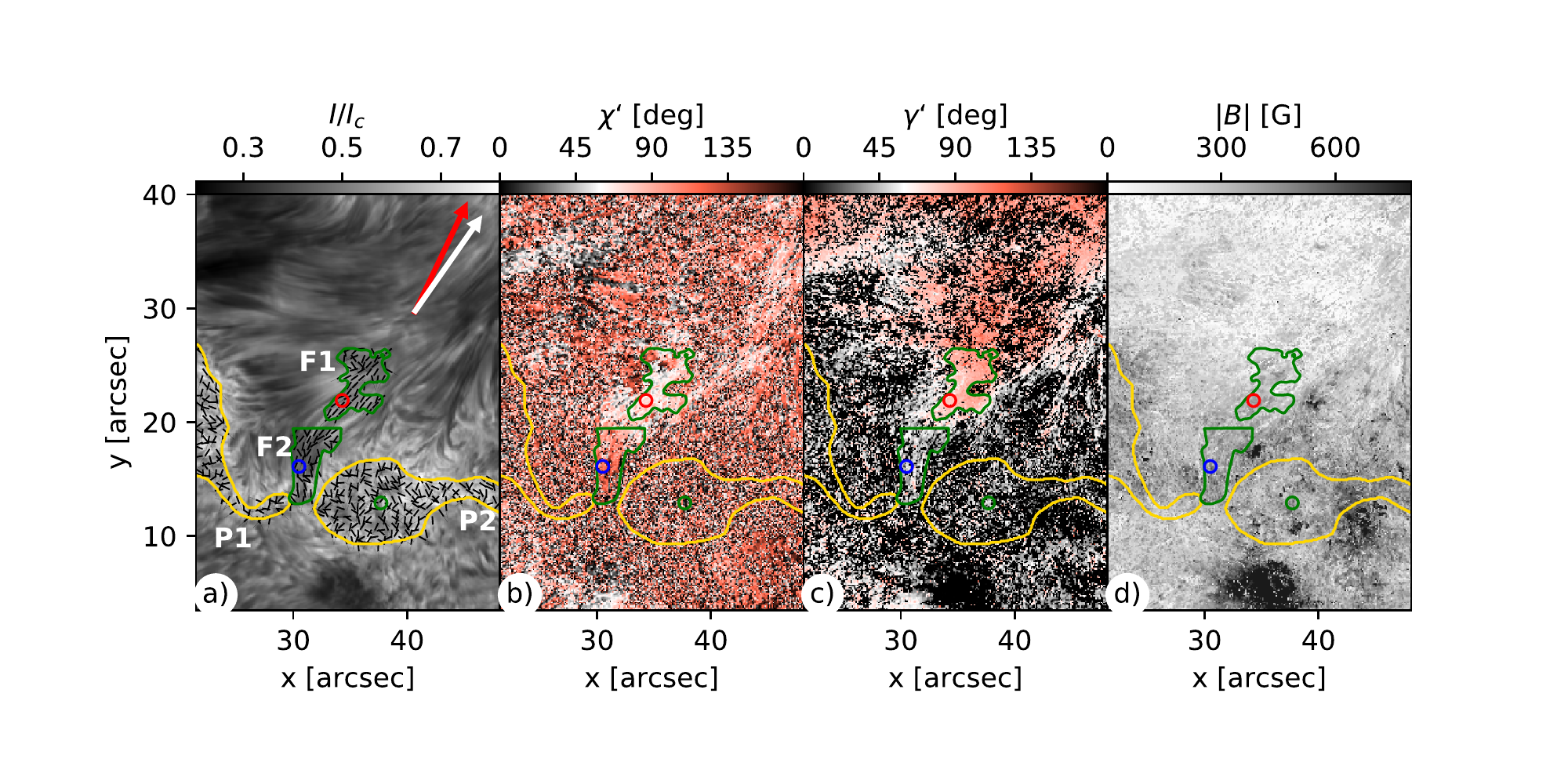}
   \caption{Magnetic field configuration. We mark the same regions as defined in Fig. \ref{fig:inversions}. Additionally we mark one pixel in each of the regions that will be used for the calculation of the uncertainties (see Fig.~\ref{fig:RF}).  a) \cair line core image with the azimuth over-plotted in our regions of interest. White arrow: Direction to disk center. Red arrow: Angle between the two pores.  b) Azimuth, c) Inclination, and d) Absolute magnetic field.} 
   \label{fig:bfield}
\end{figure*}

\begin{figure*}
   \centering
   
   \minipage{0.40\textwidth}
  \includegraphics[width=1\linewidth,trim=0cm 0cm 0cm 0cm,clip]{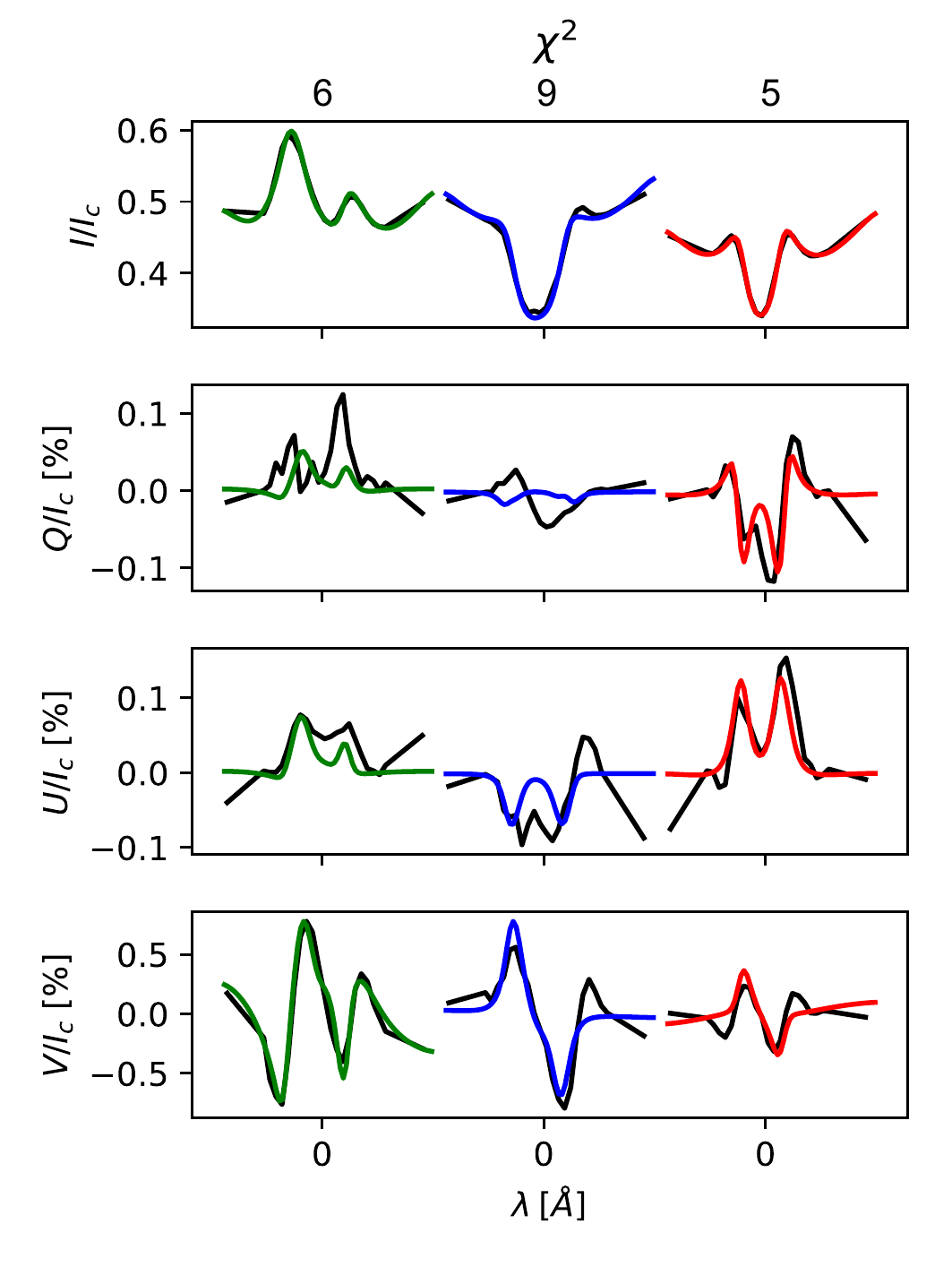}

\endminipage\hfill
\minipage{0.60\textwidth}
  \includegraphics[width=\textwidth, trim=0cm 0cm 0cm 0cm,clip]{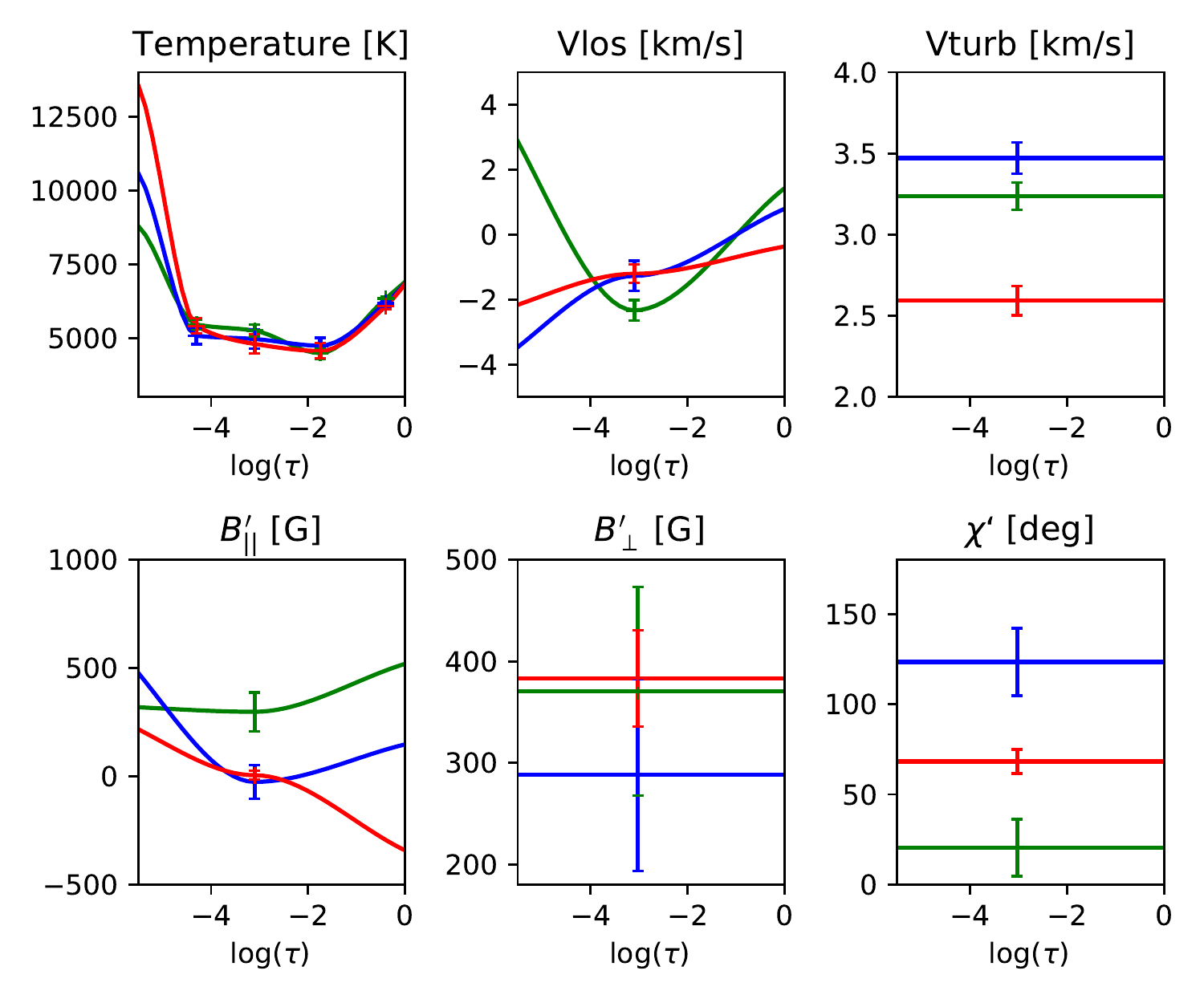}
 \endminipage  
   \caption{Left: Typical observed stokes profiles in \cair (black) and the fitted atmosphere (colored). Using Eq.~\ref{eqerr} we get a $\chi^2$ value of 6, 9, and 5 for the three profiles respectively. The pixel locations of these profiles are marked in Fig.~\ref{fig:bfield} with the same color coding.  Right: Inferred parameters displayed between $\log(\tau) = [-5.5, 0]$. The colors match the profiles on the left. The uncertainties for the same pixels based on Eq.~\ref{eqerr} are indicated by vertical bars at their respective node.}
   \label{fig:RF}
\end{figure*}

\begin{figure}
   \centering

  \includegraphics[width=1\linewidth,trim=3.5cm 0.cm 4.cm 0cm,clip]{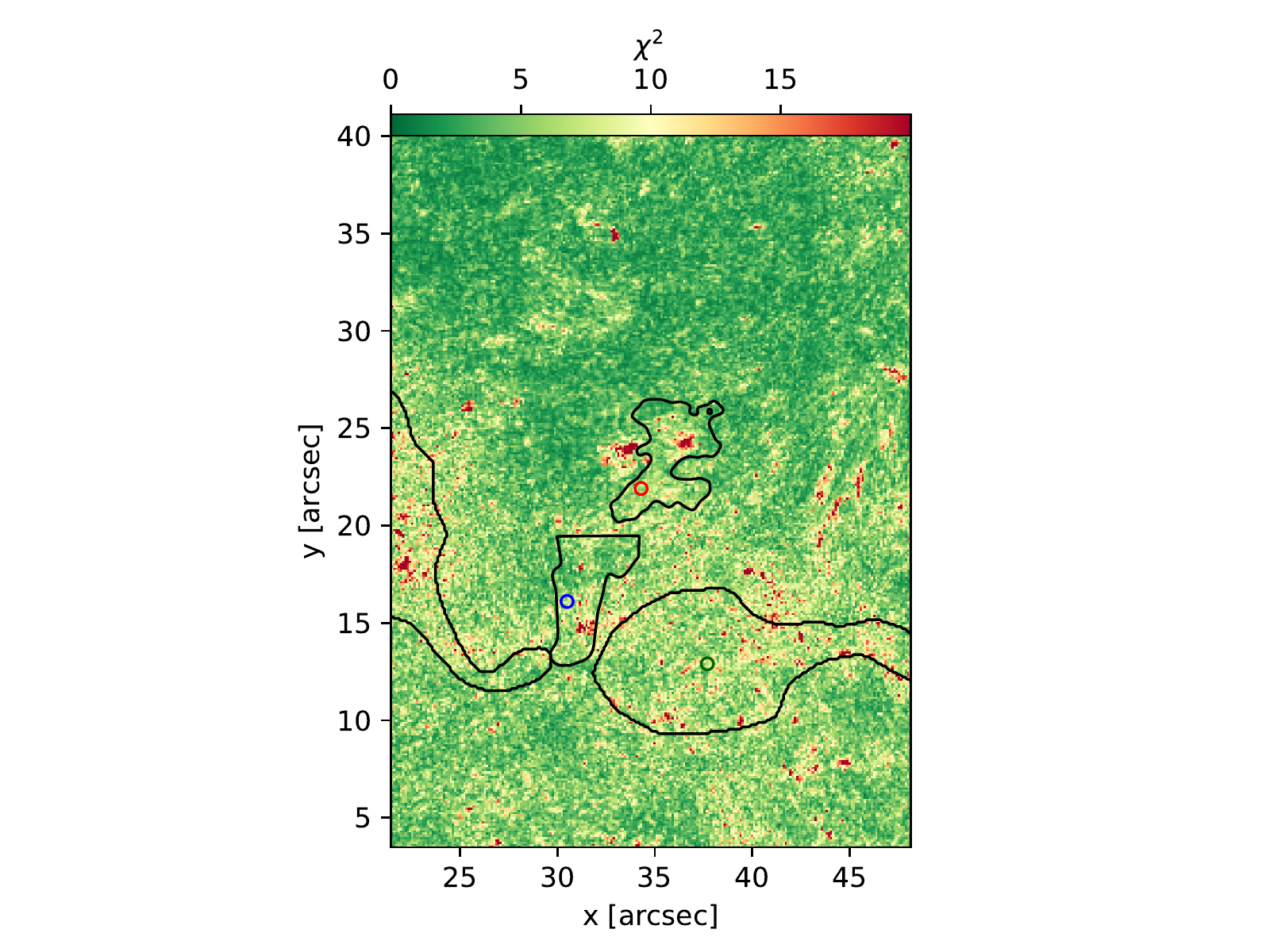}

   \caption{A $\chi^2$ map for goodness of fit based on Eq. \ref{eq:chi2}. The three selected pixels were chosen to be typical for the region that they represent. Their profiles are displayed in Fig.~\ref{fig:RF}. }
   \label{fig:chi2}
\end{figure}

From Fig.~\ref{fig:inversions}d we see that the magnetic field mostly has positive polarity in the field of view. We were able to reconstruct a longitudinal magnetic field ($B'_{||}$) throughout this FoV, but could only obtain a nonzero transverse magnetic field ($B'_{\perp}$) in certain regions. This could be due to our low-norm regularization, which favors lower values for $B'_{\perp}$. This means that higher values would only be returned if lower values did not fit the data. This was done because the transverse magnetic field is strongly affected by the noise, and if the regularization was not used we would obtain upper limits for the transverse field in all the pixels  \citep{2012MNRAS.419..153M,2019A&A...623A.178D}. With the regularization we can be sure that the obtained values are not dominated by the noise contribution. In Fig. \ref{fig:RF} (left column) the Stokes $Q$ and $U$ fits are not always good. This is common when the signal is weak and the field mostly longitudinal (as expected in the plage region). Still, there is information in these parameters and we argue that the overall results are meaningful as witnessed by the appearance of the parameter maps. Notably the value of the total magnetic field is similar in the several subregions studied.

We use $B'_{||}$ and $B'_{\perp}$ to calculate the absolute magnetic field ($|B|$) and its inclination ($\gamma'$). We show these two quantities together with the azimuthal direction of the field ($\chi'$) in Fig.~\ref{fig:bfield}. Like with $B$, the primed quantities denote the azimuth and inclination in the line-of-sight frame, whereas their unprimed counterparts represent the same quantities with respect to the solar surface.
The map of $B'_{\perp}$ shows one relatively homogeneous large region around $(x,y)=(35\arcsec,25\arcsec)$ in the area covered with fibrils. Our interpretation is that the inversions have been largely successful here. This region can be split in two based on the value of $\chi'$ in Fig.~\ref{fig:bfield}a and b. We designated these two sub-regions F1 and F2, consisting of 1334 and 1418 pixels respectively.

In the two plage regions P1 and P2, we see that the map is dotted with pixels indicating 0~G. These pixels suggest that there is a cut-off point where low values for $B_\perp$ get forced to 0 G.

From Fig.~\ref{fig:bfield}a and b, we note that $\chi'$ correlates with the signal in $B'_{\perp}$, with most of the coherent structure being found inside and near F1 and F2. The rest of the FoV shows no preferred direction. This is expected inside the plage, but in most of the fibrillar areas it is most probably due to a lack of signal.  In F1 we find a median value of $\chi' \approx 70^\circ$ and in F2 $\chi'$ seem to gradually turn from $70^\circ$ to about $130^\circ$ as we move downward in the figure.

It is not possible to disambiguate the entire field because $B_{\perp}$ could only be recovered in certain regions.  However, we can see that F1 and F2 are magnetically connected and that the magnetic field in F2 veers off into either P2 or the area between P1 and P2. Also the field in F1 is mostly aligned with the line between the two pores. The pores themselves lie along a line that is offset by only $8^\circ$ from the radius vector. We thus conclude that field lines originate from the left part of P2, flow through F2 and into F1, and reach the pore of opposite polarity outside the ROI.

The inclination map in Fig.~\ref{fig:bfield}c reflects the $B'_{\perp}$ map in Fig.~\ref{fig:inversions} since any pixel without a transverse component automatically gets interpreted as having $\gamma' = 0$ or $180^\circ$. We excluded these pixels from our analysis by selecting only those that had a $B'_\perp > 100 G$ in order to avoid the bias created by the regularization. After applying this mask, we obtained median values of the magnetic field properties for our regions of interest. Because we know that the field in P1 and P2 is close to being vertical and that the $\chi$ of F1 and F2 is close to being aligned with the radial direction, we could subtract the viewing angle to get values with respect to the local vertical, $\gamma = \gamma\lq - 37^\circ$. 

Now that we have shown that the magnetic field in F2 originates from the left side of P2, we would expect to see a similar median value of $|B|$ for the two regions.
By eye we can see a smooth-looking transition between P2 and F2 in Fig. \ref{fig:bfield}d. In the next paragraphs we discuss our measurements for the different plage and fibrillar regions.  The uncertainties for $|B|$ and $\gamma'$ were propagated from the uncertainties for $B'_\perp$ and $B'_{||}$ , which were obtained by using Eq.~\ref{eqerr} for the representative pixels.

\paragraph{Plage}
The median inclination for regions P1 and P2 is $\gamma' = 43^\circ \pm 18^\circ$ and $\gamma' = 45^\circ \pm 16^\circ$, respectively. With respect to the local vertical we get $\gamma = 7^\circ \pm 18^\circ$ for P1 and $\gamma = 8^\circ \pm 16^\circ$ for P2. These values match earlier estimates of the magnetic field inclination in plage made in the photosphere \citep[e.g.,][]{bernasconi94, Sanchez94, MartinezPillet1997}.

When looking at the total magnetic field, we find  $|B| = 440 \pm 90 \:\rm{G}$ for P1 and $|B| = 450 \pm 90 \:{\rm G}$ for P2, which is roughly twice what was reported by \cite{Carlsson2019}.
We note that changes in the detailed selection of the borders of the plage regions, even when including the tail ends of fibrils, do not affect these median values significantly.

\paragraph{Fibrils} In region F1 we find that $\gamma = 50^\circ \pm 13^\circ$ and for region F2 we find $\gamma = 9^\circ \pm 13^\circ$. The inclination of F2 is comparable to that found in plage regions P1 and P2 and mostly vertical with respect to the surface, while the inclination for F1 is more horizontal. 

In F2 we find more similarities to the plage, as this region has a median field strength of $|B| = 410 \pm 80$~G. Region F1 has a lower field strength with $|B| = 296 \pm 50$~G. This is higher than the upper limit of $|B| \sim$ 25-200 G suggested by \citet{moorogen2017}, based on the analysis of magnetohydrodynamic kink waves in chromospheric fibrils.

\noindent
 

\section{Conclusions}\label{conclusions}
We present high spatial, temporal, and spectral resolution  spectropolarimetric observations of a plage region in \cair. This SST data set allowed us to constrain the thermodynamical and magnetic properties of this region and its surroundings.  By combining high-cadence observations with the addition of novel and established post-processing techniques, we were able to reconstruct the Stokes $Q$ and $U$ profiles sufficiently well to infer $B'_\perp$ in several parts of the field of view. The high cadence required meant that CRISP could only observe \cair and the temporal binning prevented us from including \caK in our inversions. Our focus on high accuracy in the reconstruction of the magnetic field came at the cost of accuracy in the other parameters, especially in the plage region where it proved difficult to reconstruct an accurate velocity and microturbulence due to the raised core profiles.

These maps allowed us to retrieve a magnetic field map in the entirety for $B'_{||}$ and in selected regions for $B'_{\perp}$. These two maps allowed us to study the median inclination of the magnetic field in plage for the pixels where $B'_\perp$ was above 100 G. For both regions combined, this gave us a value of $\gamma = 10^\circ \pm 16^\circ$ with respect to the local vertical. This is a value that matches previous measurements made in the photosphere. 

Investigation of our $\chi'$ map showed that it could be disambiguated in one specific location of the canopy. We concluded that the magnetic field flowed from region P2 into region F2. Both of these regions have an absolute field  strength close to $450$~G ($450 \pm 90$~G in P2 and $410 \pm 80$~G in F2) and similar inclinations of $\gamma = 8^\circ \pm 16^\circ$ for P2 and $\gamma = 9^\circ \pm 13^\circ$ for F2. The fibrils in F1 have a lower field strength of $|B| = 296 \pm 50$~G and an inclination of $\gamma = 50^\circ \pm 13^\circ$.

In a parallel study, \citet{roberta2020} have studied the stratification of canopy fields using three spectral lines that provide different opacity windows in the atmosphere (\ion{Mg}{i}~5173~\AA, \ion{Na}{i}~5896~\AA\ and \ion{Ca}{ii}~8542~\AA) in combination with a spatially coupled weak-field approximation method. Although their observations did not provide a sufficiently high S/N ratio to derive the horizontal component of the field from the 8542~\AA\ line, their inferred $B_\parallel$ values in the magnetic canopy are close to 450~G in a similar target, and in agreement with our values. In our study we could not study the stratification of the field, but we could derive the full magnetic vector and also the orientation of the field along fibrils that are anchored outside of the plage target.

Our results are fundamentally limited by the evolution of the solar scene, which constrains the cadence. Therefore we believe that significant improvement would require an integral-field spectrometer allowing additional lines to be observed while keeping the cadence the same, and/or a telescope with more light-gathering power.

\begin{acknowledgements}
JdlCR is supported by grants from the Swedish Research Council (2015-03994), the Swedish National Space Agency (128/15) and the Swedish Civil Contingencies Agency (MSB). This project has received funding from the European Research Council (ERC) under the European Union's Horizon 2020 research and innovation program (SUNMAG, grant agreement 759548).
The Swedish 1- m Solar Telescope is operated on the island of La Palma by the Institute for Solar Physics of Stockholm University in the Spanish Observatorio del Roque de los Muchachos of the Instituto de Astrof\'isica de Canarias. The Institute for Solar Physics is supported by a grant for research infrastructures of national importance from the Swedish Research Council (registration number 2017-00625).
Computations were performed on resources provided by the Swedish Infrastructure for Computing (SNIC) at the PDC Centre for High Performance Computing (Beskow, PDC-HPC), at the Royal Institute of Technology in Stockholm as well as the National Supercomputer Centre (Tetralith, NSC) at Link\"oping University.
This work was supported by the Knut and Alice Wallenberg Foundation.
This research has made use of NASA's Astrophysics Data System Bibliographic Services. 
We acknowledge the community effort devoted to the development of the following open-source packages that were used in this work: numpy (numpy.org), matplotlib (matplotlib.org), astropy (astropy.org).
We thank David Buehler for a stimulating discussion on low frequency fringes and how to remove them.

\end{acknowledgements}

\bibliographystyle{aa}
\bibliography{ref}


\end{document}